\documentstyle[prl,aps,preprint,psfig]{revtex}

\tightenlines

\begin{document}
\draft 


\title{The reconstruction of Rh(001) upon oxygen adsorption}

\author{Dario Alf\`e,$^1$\cite{new} 
Stefano de Gironcoli,$^1$\cite{email} 
and Stefano Baroni$^{1,2}$\cite{email}} 
 
\address{$^1$INFM -- Istituto Nazionale di Fisica della Materia and \\
SISSA -- Scuola Internazionale di Studi Superiori ed Avanzati, via
Beirut 2-4, I-34014 Trieste, Italy\\
$^2$CECAM, 
ENSL, Aile LR5, 46 All\'ee d'Italie, F-69364 Lyon Cedex 07, France}

\date{January 9, 1998}
\maketitle

\begin{abstract} We report on a first-principles study of the
structure of O/Rh(001) at half a monolayer of oxygen coverage,
performed using density-functional theory. We find that oxygen atoms
sit at the center of the {\it black} squares of a chess-board,
$c(2\times 2)$, pattern. This structure is unstable against a rhomboid
distortion of the {\it black} squares, which shortens the distance
between an O atom and two of the four neighboring Rh atoms, while
lengthening the distance with respect to the other two. We actually
find that the surface energy is further lowered by allowing the O atom
to get off the short diagonal of the rhombus so formed. We predict
that the latter distortion is associated with an order-disorder
transition, occurring below room temperature. The above rhomboid
distortion of the square lattice may be seen as a rotation of the
empty, {\it white}, squares. Our findings are at variance with recent
claims based on STM images, according to which it is instead the {\it
black} squares which would rotate. We argue that these images are
indeed compatible with our predicted reconstruction pattern.

\vspace{0.1cm}

\end{abstract}


\narrowtext

\section{Introduction} 

The study of the physical and chemical properties of the (001) surface
of rhodium is an important step towards the ambitious goal of
understanding the catalytic properties of transition metals.
Recently, the clean Rh(001) surface has attracted experimental
\cite{Wat78,Oed88,bengley93} and theoretical
\cite{feibelman,methfessel,morrison,cho,Strnz,ChoSh} interest because
of the anomalously small interlayer relaxation at the surface, an
unexpected result which is in contrast with the general behavior of
transition-metal surfaces. One more reason of interest is due to the
behavior of the surface upon oxygen adsorption. Oxygen adsorption on
Rh (001) is known to be dissociative and to saturate at half a
monolayer, independently on the adsorption temperature. At this
coverage a $2\times 2$ reconstruction has been observed by SPA-LEED
and confirmed by STM \cite{Mercer}. The oxygen atoms sit in the
troughs formed by four first-layer rhodium atoms and fill these sites
in a c($2\times 2$) geometry. This geometry may be seen as a
chess-board whose {\it black} squares are occupied by oxygen atoms,
while the {\it white} ones are empty. Within this picture, the
reconstruction observed in Ref. \cite{Mercer} has been described as a
rotation of the {\it black} squares, resulting in a ($2 \times 2$)p4g
symmetry (see Fig. \ref{ricostruzione}a). This distortion preserves the
shape of the {\it black} squares, while the {\it white} ones become
rhomboid. A similar behavior is observed for nitrogen and carbon
adsorbed on the (001) surface of nickel, where the rotation angle of
the squares is much larger and the {\it clock} reconstruction more
evident \cite{Daum86,onuferko79,klink93,leibsle93}.

In this paper we present an {\it ab-initio} study of the oxygen
adsorption on the Rh(001) surface. In agreement with experimental
findings, we find that the preferred adsorption site is the {\it
hollow} one, at the center of the square formed by four first-layer
rhodium atoms. However, at variance with recent claims based on STM
experiments \cite{Mercer}, we find that it is the {\it white} squares
which rotate rather than the {\it black} ones (see
Fig. \ref{ricostruzione}b). In spite of this apparent discrepancy, we
argue that our predicted reconstruction pattern is indeed compatible
with experimental findings. Actually, the symmetry of our predicted
reconstruction pattern slightly differs from what appears in the
experimental STM pictures, in that ad-atoms get off the center of the
distorted squares, thus resulting in an {\it asymmetric clock}
reconstruction.
The existence of two inequivalent low-symmetry sites that oxygen atoms
can occupy in the rhombi should give rise to an order-disorder
transition whose properties, however, lie beyond the scope of the
present study.  The amplitude of the asymmetric component of the
reconstruction might be too small to be detected
experimentally. Furthermore, it is likely that the temperatures at
which these STM images have been taken lie above the critical
temperature of transition. In order to understand the driving
mechanism of the reconstruction, this is better visualized as a
rhomboid deformation of the squares filled by adsorbed oxygen
atoms. This distortion can then be rationalized by a simple chemical
model of the re-bonding of O atoms at the surface.

\section {Computational method} Our calculations are based on density
functional theory within the local-density approximation
\cite{HoKo64,KoSh65}, using Ceperley-Alder exchange-correlation (XC)
energies \cite{CeAl}. The one-particle Kohn-Sham equations are solved
self-consistently using plane-wave basis sets in a pseudopotential
scheme. Because of the well known {\it hardness} of the
norm-conserving pseudopotentials for the O and---to a lesser
extent---Rh atoms, we make use of {\it ultra-soft} pseudopotentials
\cite{Va90} which allow an accurate description of the O and Rh
valence pseudo-wave-functions with a modest basis set including plane
waves up to a kinetic-energy cutoff of 30 Ry. In the case of Rh, we
found it convenient to treat the $s$ and $p$ channels using a
norm-conserving potential, while the ultra-soft scheme is applied only
to the {\it hard} $d$ orbital \cite{stokbro,StBa,alba}. Brillouin-zone
integrations have been performed using the Gaussian-smearing
\cite{MePa} special-point \cite{MoPa} technique. We find that the
structural properties of bulk rhodium are well converged using a
first-order Gaussian smearing function \cite{MePa} of width
$\sigma=0.03 \rm ~Ry$ and $10$ special {\bf k}-points in the
irreducible wedge of the Brillouin zone (IBZ). The isolated surface is
modeled by a periodically repeated super-cell. We have used the same
super-cell for both the clean and the O-covered surfaces. For the
clean surface we have used $5$ atomic layers plus a vacuum region
corresponding to $\approx 6$ layers. For the O-covered surface, the
$5$ Rh layers are completed by one layer of O molecules on each side
of the slab: in this case the vacuum region is correspondingly reduced
to $\approx 5$ atomic layers. We have used the same Gaussian-smearing
function as in the bulk calculations with a $(12,12,2)$ Monkhorst-Pack
mesh \cite{MoPa} resulting in $21$ special {\bf k}-points in the
$1\times 1$ surface IBZ. Convergence tests performed with a value of
$\sigma$ twice as small and a correspondingly finer mesh of special
points resulted in no significant changes in total energies and
equilibrium geometries. The latter are found by allowing all the atoms
in the slab to relax until the force acting on each of them is smaller
than $0.1\times 10^{-3} {\rm Ry}/{\rm a_{0}} $.

\section {Results}

For the clean surface we find that---in agreement with other {\it
ab-initio} calculations
\cite{feibelman,methfessel,morrison,cho,Strnz,ChoSh}---the first layer
relaxes inward by $3.8 \pm 0.2~\%$ (see Table \ref{tabella}), while the second
layer is practically unrelaxed (the error is essentially a finite-size
effect, and it is estimated by repeating the calculation using thicker
slabs). As it was noted in the introduction, the value of the
first-layer relaxation reported in the experimental literature is
anomalously small (the first interlayer spacing is practically equal
to its bulk value within error bars). Recently, Cho and Scheffler
\cite{ChoSh} pointed out that a proper account of the vibrational
contribution to the surface free energy may results in a reduction of
the inward relaxation of the first layer, thus bringing theoretical
predictions in better agreement with experiments. No attempts of
estimating these vibrational effects have been done in the present
work. Our calculated values for the surface energy, $\sigma$, and work
function, $\phi$, are $\sigma \approx 1.4~\rm eV/atom$ and $\phi
\approx 5.5~\rm eV$ (see Table \ref{tabella}).

A special care must be
payed when estimating the latter, for the XC potential goes to zero
rather slowly in the vacuum region.
As the electrostatic (Hartree) potential converges much more rapidly
than the XC one to its vacuum value, it is convenient to evaluate the
vacuum level by simply neglecting the XC contributions.  As a
preliminary study of oxygen adsorption, we performed a couple of
simple {\it ab-initio} molecular dynamics simulations of an $\rm O_2$
molecule impinging onto the surface. We have tried two possible
initial conditions for the molecule, in both of which the axis of the
molecule is parallel to the surface. In the first case the projection
of the center of the molecule on the surface falls on top of a surface
atom, while in the other it falls on a bridge site. The initial
velocity of the molecule is orthogonal to the surface and its modulus
corresponds to a temperature of $300^{\rm o}$K, while the surface is
initially assumed to be at zero temperature. The initial distance of
the molecule from the surface is $\approx 6 $ a.u. which is
essentially in the vacuum. When the molecule arrives at the surface
the molecular bond breaks and the surface heats up. In order to find
the ground-state atomic configuration, we have slowly cooled down the
system, ending in the structure displayed in Fig.
\ref{nuovastruttura}: the oxygen atoms adsorb in every second hollow
site of the surface, resulting in a c$(2\times 2)$ structure, where
the squares whose centers are occupied by an ad-atom are deformed into
rhombi; the oxygens do not stay in the middle of the short diagonal of
the rhombus so formed, but move away in the orthogonal direction,
becoming essentially {\it three-fold} coordinated. As a result of the
reduced symmetry so obtained, the atomic rows of the first Rh layer
are no longer equivalent to each other: every second row is formed by
atoms which have 2 oxygen neighbors, while the others have only 1
neighbor. As a consequence the first rhodium layer results to be
buckled, the 2-fold coordinated rows leaning $\approx 0.08 $ \AA\
outwards.

This {\it asymmetric clock} reconstruction can be imagined as the
succession of two steps: in the first, the occupied, {\it black},
squares distort while the ad-atoms stay at the center of the rhombi
({\it symmetric clock} reconstruction); in the second, the ad-atom
depart from the centers of the rhombi, in order to achieve three-fold
coordination. Both the unreconstructed and {\it symmetric clock}
structures result to be saddle points whose energies are reported in
Table \ref{tabella}. We have explicitly verified that the
unreconstructed structure is stable against analogous distortions
affecting the {\it white} squares. For the sake of completeness, in
Table \ref{tabella} we also report the relative surface energies
corresponding to two other adsorption sites (see Fig. \ref{siti}),
which are however much larger than the energy differences between the
unreconstructed {\it hollow} structure and the corresponding {\it
clock} reconstructions.

The work function is larger at coverage $\Theta = {1 \over 2}$ than
for the clean surface (see Table \ref{tabella}), thus indicating that
electrons tend to transfer from the Rh substrate to the oxygen layer,
so as to increase the surface dipole. In Fig. \ref{coverage} we
display the adsorption energy, $E_{ad}$, as a function of coverage,
$\Theta$, defined as: $E_{ad}(\Theta) = \sigma(\Theta) - \sigma(0)
-{\Theta\over 2} E_{\rm O_2} $, where $\sigma(\Theta)$ is the surface
energy of the oxygen-covered system, and $E_{\rm O_2}$ is the energy
of the isolated $\rm O_2$ molecule. Coverage $\Theta=1$ is realized by
filling all the hollow sites; for $\Theta={3\over 4}$ and
$\Theta={1\over 4}$, 3 and 1 of the 4 possible hollow sites in a $2
\times 2$ substrate super-cell have been occupied. The minimum energy
corresponds to a coverage in the range $ {1\over 2} \lesssim \Theta
\lesssim 1$. In this range, however, the curve is very flat and the
variation of the adsorption energy ($\approx 100 \rm ~meV$) is smaller
than the expected accuracy of the terms which enter its
definition. All we can safely predict is that the stable coverage
state lies somewhere in between $\Theta = {1\over 2}$ and $\Theta
=1$. The value of the most stable coverage is determined by a
trade-off between the adsorption energy of an isolated oxygen
molecule, which tends to favor a high coverage, and the oxygen-oxygen
electrostatic repulsion due to charge transfer, which becomes more
effective when the average $\rm O-O$ distance becomes smaller than
some typical screening length, and which tends instead to favor a low
coverage.

\section{discussion}

From Fig. \ref{nuovastruttura} it is evident that the oxygen
sub-lattice forms a {\it zig-zag} arrangement which is not observed in the
experiments \cite{Mercer}. However, there are two equivalent threefold
sites for each cell occupied by an oxygen ad-atom, one on each side of
the bridge. Neglecting the interactions between different adsorption
sites, each of them is therefore two-fold degenerate. We postulate
that the system undergoes an order-disorder transition at some
critical temperature which depends on the magnitude of the
adsorbate-adsorbate interactions, and that this temperature is lower
than that at which the STM images have been taken. The fact that no
disorder in the O ad-layer appears from these images indicates that
the frequency of oxygen barrier crossing is much larger than the
scanning frequency of the STM. This is compatible with our estimate of
the barrier height ($\approx 30$ meV, as obtained by comparing the
energies of the {\it symmetric clock} and the {\it asymmetric clock}
structures) which is of the order of the room temperature.

The STM pictures reported in Ref. \cite{Mercer} are such that only Rh
or O atoms are visible in turn in a same picture. Because of this, it
is difficult to judge whether the squares that undergo a rhomboid
distortion are those filled by an ad-atom, or they are instead empty.
What appears to be rather evident is that the distorted squares
display a depletion with respect to the undistorted ones. In analogy
with a similar reconstruction which is observed to occur in N/Ni(001)
and which is much better characterized \cite{wenzel}, the authors
of Ref. \cite{Mercer} conclude that O atoms are thrust down the first
Rh layer in the middle of the undistorted squares. We have simulated
STM pictures using the Tersoff-Hamann model \cite{TeHa}, but we have
not been able to eliminate the images of the O atoms from any of our
simulated pictures. In all of them, O atoms appear as protrusions in
the middle of the distorted squares, in qualitative agreement with the
experimental observation that undistorted squares are depleted with
respect to the distorted ones.

In Fig. \ref{proj} we display the surface-projected densities of
states (SDOS) of the clean and of the oxygen-covered Rh(001) surface,
along with their decompositions into various atomic-like
contributions. The SDOS of the clean surface differs from its bulk
counterpart mainly because of its more pronounced narrowness due to
the lower coordination of surface atoms \cite{Strnz,StBa}, and it is
almost entirely determined by its $d$-like component (see
Fig. \ref{proj}a). In the bulk FCC structure the three $d_{xy},
d_{xz}$ and $d_{yz}$ are degenerate, and so are $d_{3z^2-r^2}$ and
$d_{x^2-y^2}$. This degeneracy is partially lifted at the surface
because the surface atoms lose the neighbors in the direction
orthogonal to the surface. For the (001) surface, the $d_{xz}$ and
$d_{yz}$ orbitals are still equivalent by symmetry and, hence,
degenerate. The position of the oxygen atomic $p$-level is around 3.1
eV below the Fermi energy of the oxygen-covered surface. Inspection of
Fig. \ref{proj}d shows that the O$_{p_x}$ = O$_{p_y}$ level gives rise
to a bonding and an anti-bonding main peaks, respectively below
($-5.8~\rm eV$) and above ($+1~\rm eV$) the Fermi level. Upon oxygen
adsorption, first-layer Rh atoms become locally inequivalent according
to whether the neighboring oxygen atoms are aligned along the $x$ or
$y$ directions (see Fig. \ref{proj}e). The projected densities of
states (PDOS) plotted in Fig. \ref{proj}d refer to those Rh atoms which have O
neighbors aligned along the $y$ direction (the PDOS of the other Rh
atoms can be obtained by simply exchanging $x$ with $y$). It is also
easy to recognize that the same bonding and anti-bonding features
occur in the $d_{yz}$ band whose atomic orbitals have lobes oriented
towards the adsorbed oxygen, while the $d_{xz}$ band remains similar
to that of the clean surface (see Fig. \ref{proj}c) because in that direction
the surface sites are empty. The O$_{p_x}$ orbitals make bonds with
the $d_{xz}$ orbitals of the Rh atoms along the $x$ direction, while
the O$_{p_y}$ orbitals hybridize with the $d_{yz}$ orbitals of the Rh
atoms along the $y$ direction. We find that the c($2 \times 2$)
structure is not the most stable one and that a ($2 \times 2$)p4g
reconstruction occurs which shortens and strengthens the bonds O$-$Rh
along one direction, while lengthening them in the perpendicular
direction, as illustrated in Fig \ref{proj}e. The amplitude of the
distortion depends rather sensitively on the lattice parameter: using
our calculated lattice parameter, we estimate the distortion to be
$\approx 4\%$; if one uses a lattice parameter $1\%$ larger, the
amplitude of the distortion also is increased, reaching a value
$\approx 6\%$. The opposite occurs if the lattice parameter is
reduced. The tendency of the system to strengthen two of the four
rhodium bonds at the expenses of the other two results in a net
lowering of the surface energy of $\approx 3$ meV/atom, thus
stabilizing the ($2 \times 2$)p4g structure. In the {\it asymmetric clock}
more stable structure the amount of the distortion is $\approx 11\%$
(using our calculated lattice parameter) and the surface energy gain
is $\approx$ 30 meV/atom. This shows that the reconstruction occurs
because the optimal O$-$Rh bond-length is shorter than that realized
in the ideal geometry. The {\it chemical} contribution to the energy
lowering which determines the distortion is illustrated in
Fig. \ref{proj}f which shows the differences between the Rh$_d$ and
O$_p$ PDOS after and before the reconstruction. In both cases, we
notice a push of states towards lower energies.

\section{Conclusions}
Our {\it ab-initio} results confirm the experimental evidence of the
dissociative character of the oxygen adsorption on Rh (001) and that
the favored adsorption site is the {\it fourfold} one. A ($2 \times
2$)p4g reconstruction of the surface is also predicted, in agreement
with SPA-LEED data. At variance with claims based on recent STM work,
we find that this reconstruction is due to a rhomboid distortion of
the squares formed by first-layer Rh atoms which have an O atom
adsorbed in the middle, and that this structure is unstable with
respect to a departure of the ad-atoms from the centers of the
resulting rhombi. We argue that the experimental evidence upon which
these claims are founded is rather weak and we suggest therefore that
further experimental work is needed to fully characterize the
reconstruction of this surface. Furthermore, we find evidence that an
order-disorder transition should occur at some temperature below that
at which the STM pictures have been taken. Our findings are
substantiated by a simple chemical model of the mechanisms responsible
for the reconstruction.

\vskip 20pt Our calculations were performed on the SISSA IBM-SP2 and
CINECA-INFM Cray-T3D/E parallel machines in Trieste and Bologna
respectively, using the parallel version of the {\tt PWSCF}
code. Access to the Cray machines has been granted within the {\it
Iniziativa Trasversale Calcolo Parallelo} of the INFM.

\begin{table}
\caption{ Structural data for the three oxygenated structures
investigated (see Fig. \ref{siti}) and for the clean surface. The
coverage is $\Theta = {1 \over 2}$. $d_0$ is the bulk lattice spacing,
$d_{01}$ is the distance between the oxygen atoms and the first
rhodium layer, $d_{12}$ the distance between the first and the second
layer. For the asymmetric clock reconstruction 
the two given numbers refer to the
two inequivalent first layer rhodium atoms (see
Fig. \ref{nuovastruttura}). $\delta$ is the amplitude of the movement
of the first-layer rhodium atoms upon distortion (see
Fig. \ref{siti}(b)), $\phi$ is the work function and $\sigma$ is the
surface energy.}
\label{tabella}
\begin{tabular}{lcccccc}
& $\delta/d_0$ & $d_{01}$ & $d_{12}/d_0 $ &
 $\phi$ & $E-E_{Hollow}$ & $\sigma$ \\ 
unit & $\phantom{spa}$ \% $\phantom{spa}$ & \AA & \% & (eV) & (eV/at)
& (eV/et)\\ 
\tableline
Clean & & & $-3.8\%$ & $5.5$ & & $1.36$ \\ 
Expt. & & & $-1.2\pm 1.6\%\tablenotemark[1] $ 
& $5.0\tablenotemark[2]$ & & $ 1.27 \tablenotemark[3] $ \\
Top & & $1.81$ & $+3$ & $7.5$ & $+1.5$ & \\ 
Bridge & & $1.33$ & $+0.5$ & $6.8$ & $+0.3$ & \\ 
Hollow &  & $1.02$ & $+0.5$ & $6.2$ & $\phantom{+}0.0$ & \\ 
S. clock & $\approx \phantom{1}4\phantom{^a} $  & $1.02$ & $+0.5$ & 
$6.2$ & $ -0.003 $ &\\ 
A. clock & $\approx 11\phantom{^a} $  & $0.98\div 1.06\phantom{^a}$ 
& $+0.5\div 0.1\phantom{^a}$ & $6.1$ 
& $ -0.030 $ &\\ 
\end{tabular}
\tablenotetext[1]{From Ref. \cite{bengley93}}
\tablenotetext[2]{From Ref. \cite{handbook}}
\tablenotetext[3]{From Ref. \cite{mezey82}}
\tablenotetext[4]{From Ref. \cite{Mercer}}
\tablenotetext[5]{From Ref. \cite{Oed88}}
\end{table}

\newpage
\begin{figure}
\caption{Two possible {\it clock} reconstructions of the O/Rh(001)
surface, resulting in a $(2\times 2)p4g$ structure. In the {\it black}
reconstruction (a) the squares with an O atom in the middle rotate,
while the empty ones distort to rhombi; in the {\it white} reconstruction (b),
the opposite occurs.}
\label{ricostruzione}
\end{figure}		
\begin{figure}
\caption{Equilibrium structure of O/Rh(001) as obtained by our
simulated-annealing procedure. The thick line indicates the unit cell.
The thin line indicates the rhombus and its shorter diagonal. The
oxygen atoms are alternatively shifted orthogonally with respect to
the shorter diagonal. The atoms of the first surface layer (depicted
with a brighter tone) are $\approx 0.08 $ \AA ~ higher then the
others.}\label{nuovastruttura}
\end{figure}
\begin{figure}
\caption{Sketch of the various O adsorption sites of the Rh(001)
surfaces considered in this work. Small dark circles: O atoms. Large
lighter circles: first-layer Rh atoms. Large darker circles:
second-layer Rh atoms.}\label{siti}
\end{figure}
\begin{figure}
\caption{Oxygen adsorption energy, $E_{ad}$ (eV), as function of
coverage, $\Theta$.}\label{coverage}
\end{figure}
\begin{figure}
\caption{ {\it a}) Surface projected density of states for the clean
Rh(001) surface; {\it b}) Surface projected density of states for the
the oxygen covered one Rh(001) surface; {\it c}) Projection of the
SDOS onto the $d_{xz} = d_{yz}$ first-layer rhodium orbitals for the
clean surface; {\it d}) Projection of the SDOS onto the $p_x= p_y$
oxygen orbitals and of $d_{xz}$ and $d_{yz}$ first-layer rhodium
orbitals for oxygenated, unreconstructed, surface; {\it e}) Sketch of
the ($2 \times 2$)p4g, {\it clock}, reconstruction.  The amplitude of
the deformation is exaggerated for clarity; {\it f}) Differences
between the O$_p-$ and Rh$_d-$projected density of states in the
reconstructed and unreconstructed structures.  }\label{proj}
\end{figure}

\newpage
\centerline{
{\psfig{figure=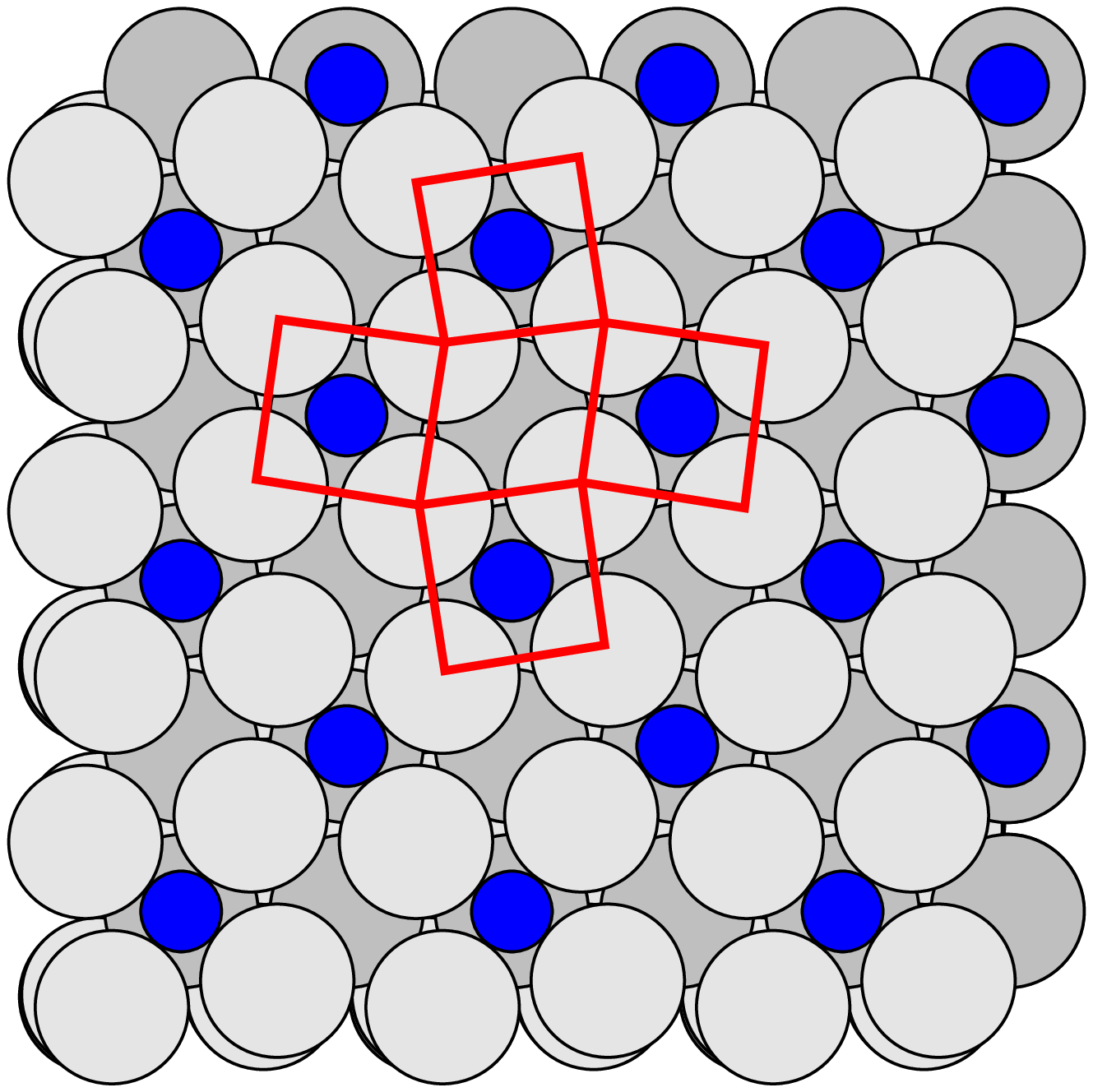,height=3.0in}}
{\psfig{figure=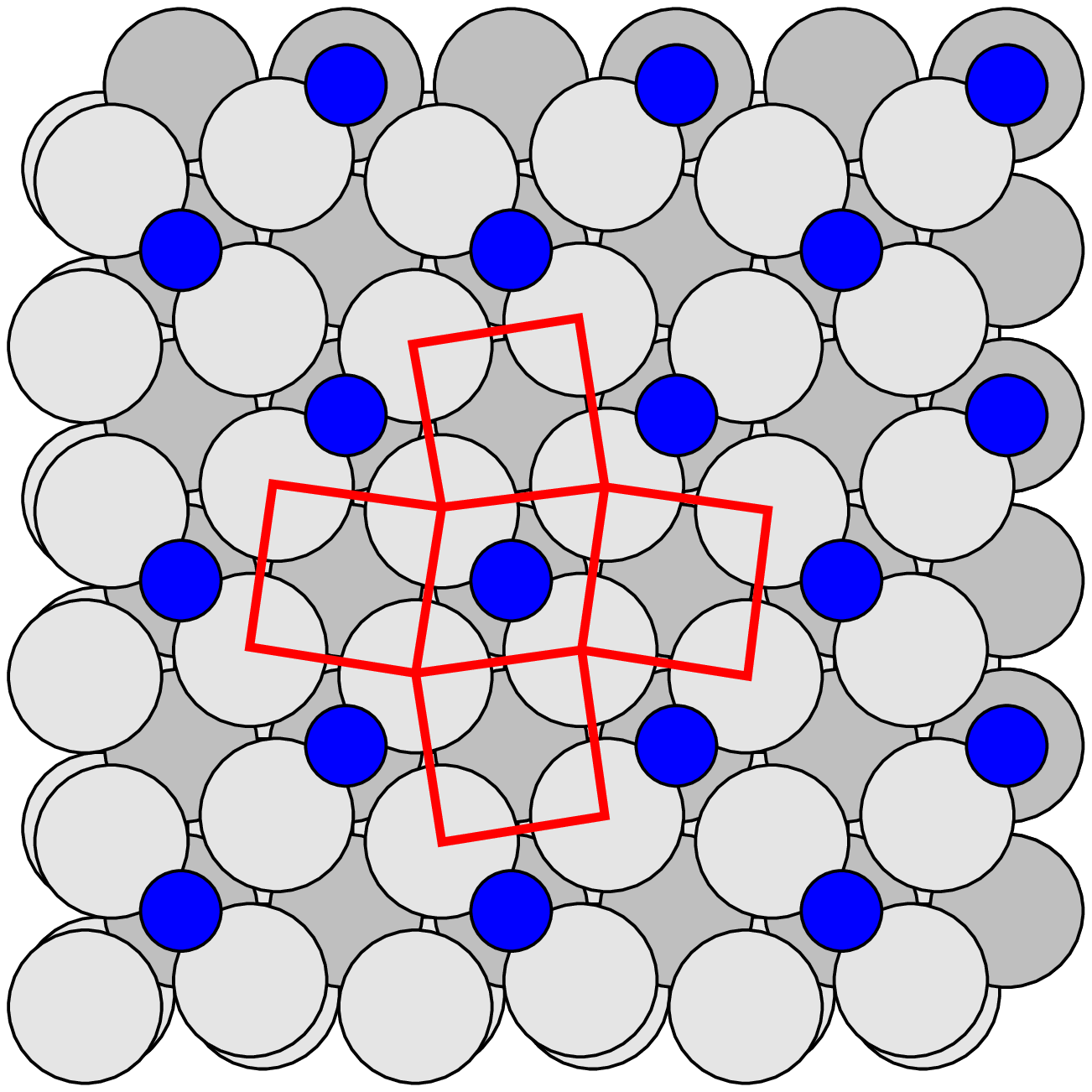,height=3.0in}}
}
\newpage
\centerline{\psfig{figure=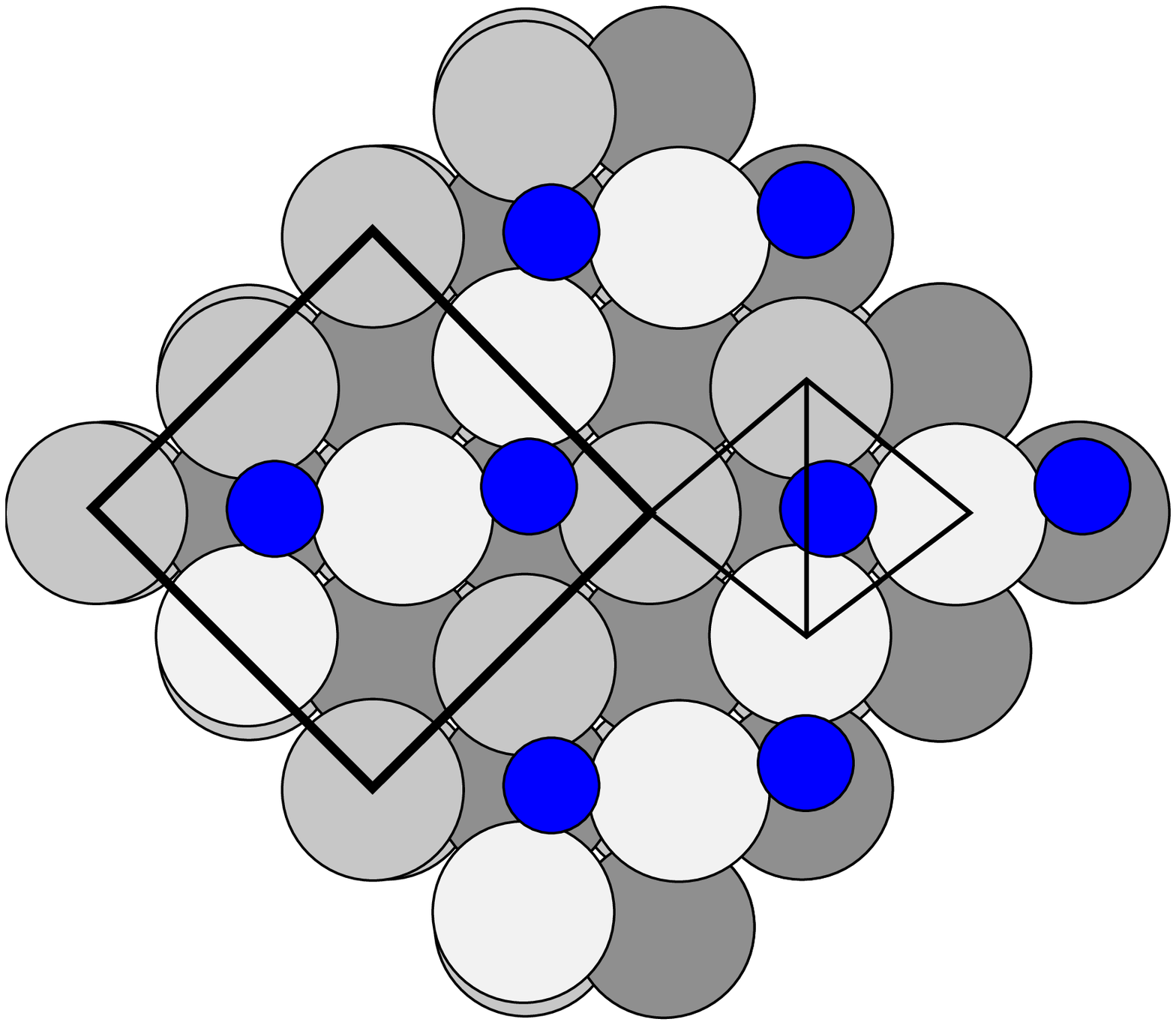,height=3in}} 
\newpage
\centerline{\psfig{figure=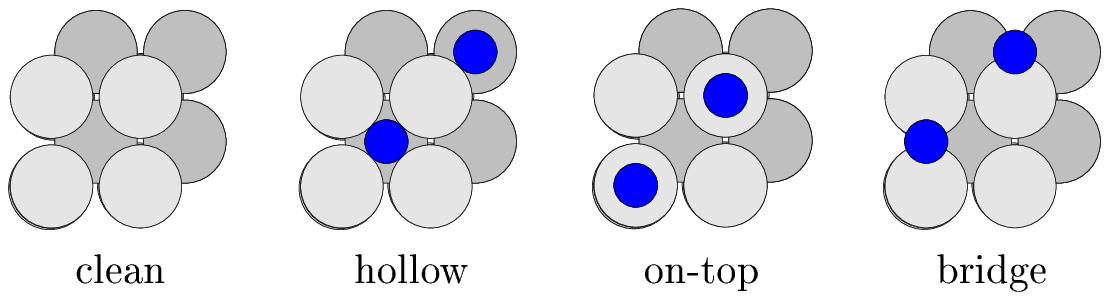,height=2in}} 
\newpage
\centerline{\psfig{figure=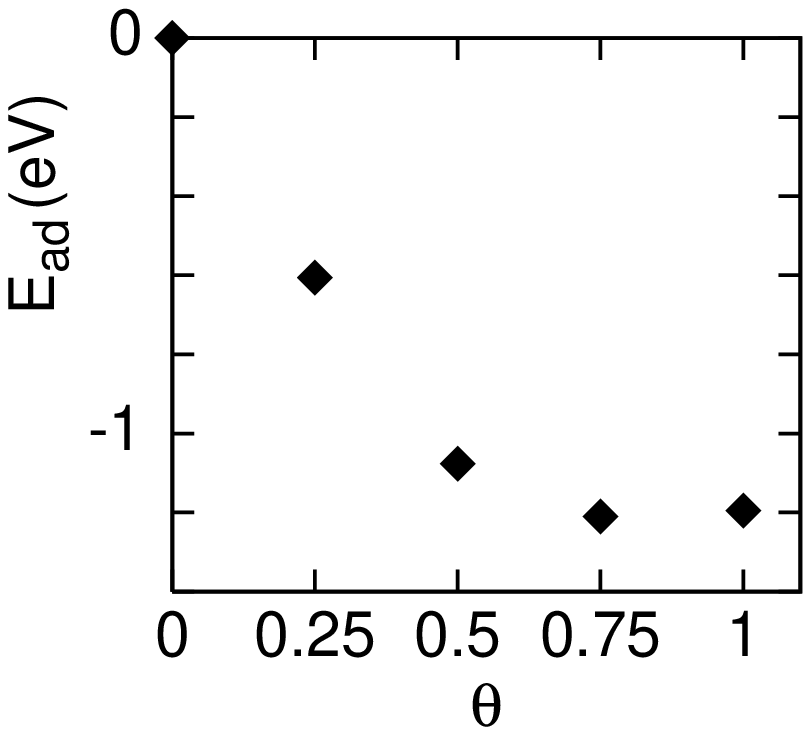,width=3.8in}} 
\newpage
\centerline{\psfig{figure=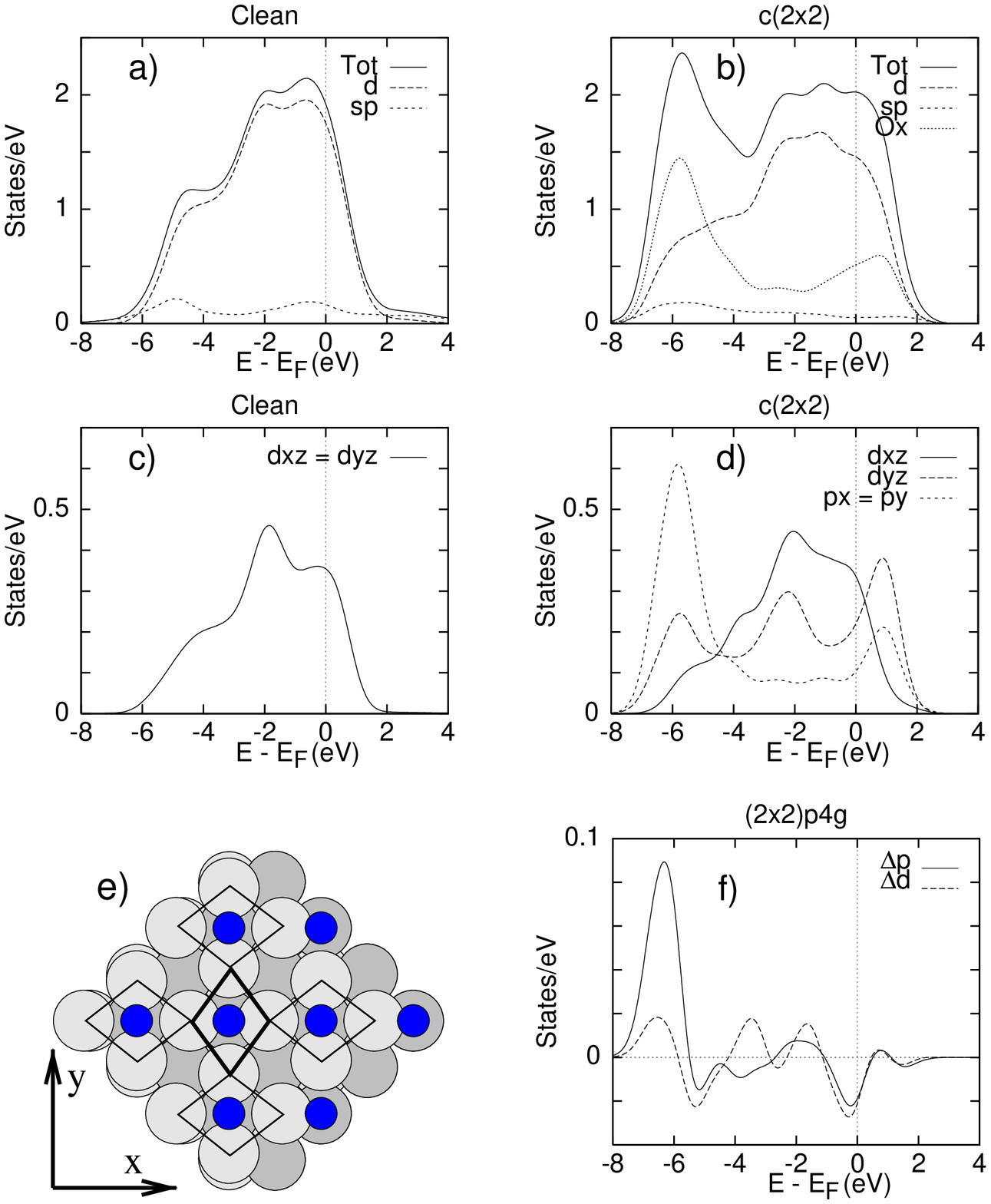,width=6in}}

\end{document}